\documentclass{article}
\usepackage{latexsym, amssymb, enumerate, amsmath}
\usepackage{graphicx}
\usepackage{amsfonts}
\usepackage{scalefnt}
\usepackage{amsmath}
\usepackage{amssymb}

\newtheorem{thm}{Theorem}[section]
\newtheorem{prop}[thm]{Proposition}
\newtheorem{lem}[thm]{Lemma}
\newtheorem{cor}[thm]{Corollary}
\newtheorem{defn}[thm]{Definition}

\begin{document}

\title{An Upper Bound for Signal Transmission Error Probability in Hyperbolic Spaces}
\author{Edson Agustini \thanks{Faculdade de Matem\'{a}tica - Universidade Federal de
Uberl\^{a}ndia - UFU - MG, Brazil, (agustini@ufu.br).}
\and Sueli I. R. Costa \thanks{Instituto de Matem\'{a}tica - Universidade Estadual
de Campinas - UNICAMP - SP, Brazil, (sueli@ime.unicamp.br).}}
\maketitle

\begin{abstract}

We introduce and discuss the concept of Gaussian probability density function
(pdf) for the n-dimensional hyperbolic space which has been proposed as an
environment for coding and decoding signals. An upper bound for the error
probability of signal transmission associated with the hyperbolic distance is
established. The pdf and the upper bound were developed using Poincar\'{e}
models for the hyperbolic spaces.

\end{abstract}

\pagestyle{myheadings} \markboth{An Upper Bound for Signal Transmission
Error Probability in Hyperbolic Spaces}{Agustini, E. and Costa, S. I. R.}

\bigskip


\section{Introduction}

\label{intro}

This paper is devoted to the introduction of a \textit{probability density
function} (\textit{pdf}) in the $n$-dimensional hyperbolic space
$\mathbb{H}^{n},$ \cite{Beardon}. The study was motivated by the diversity on
sets of points in $\mathbb{H}^{n}$ with interesting symmetry properties which
can be used in Coding Theory but which find no analogies in the Euclidean
space $\mathbb{R}^{n}.$ More specifically, in the hyperbolic space there is a
much greater possibility of uniform point allocations giving rise to regular
tilings. Such sets of points admit labeling by groups of hyperbolic symmetries
and are an extension of the well-known geometrically uniform codes,
\cite{Forney}. Concerning minimum distance there is also some advantage since
a set of $M$ equidistant points on a hyperbolic circle (hyperbolic $M$-PSK)
are farther apart then their Euclidean counterpart. On the other hand, bounds
for signal transmission error probability are usually harder to establish
since there is no compactible global vectorial structure in $\mathbb{H}^{n}.$

In some cases it may be quite natural to consider the hyperbolic distance
rather than the usual Euclidean distance. As shown in Section
\ref{Sec Gauss Dim 1},\ the lognormal \textit{pdf}, which models various
situations, can be viewed as a (symmetric) Gaussian \textit{pdf} when
described in terms of the hyperbolic distance. On the other hand, hyperbolic
Gaussian \textit{pdf's} in dimension $n,$ $n\geq2,$ can be viewed as Euclidean
\textit{pdf's} which are a combination of a lognormal \textit{pdf} (in one
direction) with a Gaussian \textit{pdf} (in the orthogonal complement of this
direction). Considering it in the hyperbolic environment has the advantage of
homogeneity since we now have symmetry. The approach chosen for this paper was
to establish the proper definition of a Gaussian \textit{pdf} when the
hyperbolic distance is considered and then deduce bounds for the associated
error probability which are analogous to the ones associated to the usual
(Euclidean) Gaussian \textit{pdf's}.

Parallel to the study of sets of points that can be used in coding systems,
there is the question of determining the interference type (noise) affecting
the transmission channel of a communication system. The hyperbolic structure
may more appropriately model certain situations where the transmission of
signals favors a specific direction in the space.

For signals transmitted in the Euclidean space $\mathbb{R}^{n},$ the main
noise type usually considered is the one associated to a Gaussian \textit{pdf}
(Additive White Gaussian Noise - AWGN channel). The analogous \textit{pdf}
introduced here for modeling the hyperbolic noise is based on the equivalent
properties of the Euclidean Gaussian \textit{pdf}.

In \cite{Agustini} we built codes starting from constellations of points in
the hyperbolic plane with some of those sets having equivalents in
$\mathbb{R}^{n}$ and others showing the peculiarities of hyperbolic geometry.
However, to perform comparative performance analyses among hyperbolic
constellations it is important to set bounds for error probability. The bounds
we deduce here include those corresponding to Bhattacharyya bounds,
\cite{Biglieri}, in the Euclidean case. The establishment of such bounds in
the hyperbolic case in terms of the number and the distance of neighboring
points from a given point constitutes the central result of this paper
(Theorem \ref{TeoremaEstimador}). In \cite{ISITAustralia} we have presented a
summary of some results of this paper.

Related matters to the subject presented here include \cite{EduardoArtigo},
which compare distances in signal constellations in the two-dimensional case
(hyperbolic and Euclidean planes). The papers \cite{Kitaev}, \cite{Bombin} and
\cite{Clarice} are about quantum codes on compact surfaces of genus $g\leq{2}%
$, which are surfaces obtained of hyperbolic plane quotiented by Fuchsian Groups.

It is interesting to note a growing interest in approaching hyperbolic
structures in recent papers like \cite{Banuelos} on Brownian motions
conditioned to harmonic functions in hyperbolic environment. It is also worthy
to point out the natural way that the hyperbolic metric appears in statistical
models as introduced by R. Rao, \cite{Atkinson}, since the distance between
two Euclidean Gaussian \textit{pdf's,} when measured by the Fisher information
metric, is hyperbolic, \cite{CostaSantos}.

This paper is organized as follows. In Section \ref{Sec Prob Limit Rn} we
introduce the Gaussian \textit{pdf} and the Bhattacharyya upper bound in the
Euclidean space $\mathbb{R}^{n}$ (Proposition \ref{PropLimitSupGaussRn}),
using an approach for error probability bounds in transmission channels which
allows similar development in the hyperbolic space $\mathbb{H}^{n}.$ In
Section \ref{Sec Gauss Dim 1} we introduce the concept the Gaussian
\textit{pdf} in the one-dimensional hyperbolic space $\mathbb{H}$ and show
that it is equivalent to the lognormal \textit{pdf}. Section
\ref{Sec Gauss Lim 2} is devoted to the development of Gaussian \textit{pdf}
in the $n$-dimensional hyperbolic space $\mathbb{H}^{n},$ $n\geq2,$ and its
peculiarities. In Section \ref{Sec Hyp Limit} we develop an upper bound for
the error probability in transmission channels of $\mathbb{H}^{n}$ (Theorem
\ref{TeoremaEstimador}). Conclusions and perspectives are drawn in Section
\ref{Sec Conclud}.

\section{\label{Sec Prob Limit Rn}Error Probability in Gaussian Transmission
Channels}

One of the concerns in coding theory is to find means to determine or to limit
the error probability associated with code words (points). For transmission
channels for signals in $\mathbb{R}^{n}$ with interferences of the type
\textit{AWGN}\footnote{Additive White Gaussian Noise - more common type of
interference considered in $\mathbb{R}^{n}.$}, an estimation of the error
probability can be made through bounds, for example, the \textit{Bhattacharyya
bound}, \cite{Biglieri}. This bound can be applied to discrete sets of signals
(points) in $\mathbb{R}^{n},$ for any $n.$

Our aim in this section is recall concepts and notations on signal
constellations and transmition error probability, to present the Proposition
\ref{PropLimitSupGaussRn} and its corollaries which will be extended to the
hyperbolic space in the next sections. Proposition \ref{PropLimitSupGaussRn}
establishes an upper bound for the error probability associated to
\textit{equally likely to be transmitted signals} (\textit{elts}) in the
Euclidean space that has the Bhattacharyya bound as particular case.

We begin introducing the concepts of \textit{geometrically uniform code} and
\textit{Voronoi (or Dirichlet) region, }for general metric spaces.

\begin{defn}
Let $\left(  \mathbb{M},d\right)  $ be a metric space, $\mathcal{C}$ a set of
points in $\mathbb{M}$. A set $\mathcal{C}$ is said a \emph{geometrically
uniform code} if $\mathcal{S}$ acts transitively in $\mathcal{C},$ that is, if
for any two points $p$ and $q$ in $\mathcal{C},$ the exists an isometry ( i.e.
a map which preserves distance) $\varphi$ in $\mathcal{S}$ such that
$\varphi\left(  p\right)  =q.$
\end{defn}

\begin{defn}
Let $\mathcal{C}$ be a set of points in a metric space $\left(  \mathbb{M}%
,d\right)  $ and $p\in\mathcal{C}.$ The set
\[
V_{p}=\left\{  r\in\mathbb{M}:d\left(  p,r\right)  \leq d\left(  r,q\right)
,\text{ }\forall q\in\mathcal{C}\right\}
\]
is denoted \emph{Voronoi (or Dirichlet) region} of $p$ in $\mathbb{M}.$
\end{defn}

We consider in this section a set $\mathcal{S}$ of $m$ \textit{elts} in
$\mathbb{R}^{n}.$ The interference (noise) $\mathbf{n}$ of the \textit{AWGN}
type acts in an additive way in $\mathbb{R}^{n},$ that is, when the signal
$\mathbf{s}_{k}$ is sent, we receive the signal%
\[
\mathbf{r(s}_{k})=\mathbf{s}_{k}+\mathbf{n(s}_{k})
\]
where $\mathbf{n(s}_{k})\in\mathbb{R}^{n}$ is determined by a Gaussian
\textit{pdf} with mean $\mathbf{0}$ and variance $\sigma^{2}$. (\footnote{The
random variable defined at the instant $t,$ $\mathbf{n}\equiv n\left(
t\right)  ,$ is a sample function of a Gaussian random process with mean
$\mathbf{0}$ and variance $\sigma^{2}.$})

If $\mathbf{r(s}_{k})$ is a signal belonging to the Voronoi region $V_{k}$ of
the signal $\mathbf{s}_{k},$ it should be natural to choose $\mathbf{s}_{k}$
as the correct decoding. Of course, if the interference $\mathbf{n(s}_{k})$ is
such that the received signal, $\mathbf{r(s}_{k}),$ is out of $V_{k},$ we have
a decoding error, since $\mathbf{s}_{k}$ won't be chosen. Hence, the correct
decoding depends on the distance among the transmitted signals, which, in its
turn, also depends on the amount of energy $E$ available in the communication system.

The \textit{Gaussian pdf} associated to $\mathbf{s}_{k}$ in $\mathbb{R}^{n}$
for \textit{AWGN} channels is given by
\[%
\begin{array}
[c]{cccc}%
g_{\mathbf{s}_{k}}: & \mathbb{R}^{n} & \longrightarrow & \mathbb{R}^{+}\\
& \mathbf{x} & \longmapsto & \dfrac{1}{\sqrt{\left(  2\pi\sigma^{2}\right)
^{n}}}\exp\left(  -\dfrac{\left\vert \mathbf{x}-\mathbf{s}_{k}\right\vert
^{2}}{2\sigma^{2}}\right)
\end{array}
\]
where $\left\vert .\right\vert $ is the usual euclidean norm in $\mathbb{R}%
^{n},$ $\sigma^{2}$ is the variance and $\mathbf{s}_{k}$ is the mean.

Therefore, the correct decoding probability is given by:%
\[
\mathcal{P}_{c}\left(  \mathbf{s}_{k}\right)  =\dfrac{1}{\sqrt{\left(
2\pi\sigma^{2}\right)  ^{n}}}\int_{V_{k}}\exp\left(  -\dfrac{\left\vert
\mathbf{x}-\mathbf{s}_{k}\right\vert ^{2}}{2\sigma^{2}}\right)  dV_{\mathbb{R}%
^{n}},
\]
where $V_{k}$ is the Voronoi region of $\mathbf{s}_{k}$ and $dV_{\mathbb{R}%
^{n}}$ is the Cartesian volume element in $\mathbb{R}^{n}.$

Since the error probability in decoding is:%
\[
\mathcal{P}_{e}\left(  \mathbf{s}_{k}\right)  =1-\mathcal{P}_{c}\left(
\mathbf{s}_{k}\right)  ,
\]
and the $m$ signals $\mathbf{s}_{k}$ are \textit{elts}, the mean error
probability $\mathcal{P}_{e}$ associated to $\mathcal{S}$ is given by
\begin{equation}
\mathcal{P}_{e}=\dfrac{1}{m}%
{\textstyle\sum\limits_{k=1}^{m}}
\dfrac{1}{\sqrt{\left(  2\pi\sigma^{2}\right)  ^{n}}}\int_{\mathbb{R}%
^{n}-V_{k}}\exp\left(  -\dfrac{\left\vert \mathbf{x}-\mathbf{s}_{k}\right\vert
^{2}}{2\sigma^{2}}\right)  dV_{\mathbb{R}^{n}}.\label{ProbErroGaussRn}%
\end{equation}

\textsc{Figure 1} shows, in perspective, the central part of the graph of
$g_{\mathbf{s}_{8}}$ in a set of 8 signals uniformly distributed on a circle
($8$\textit{-PSK}\footnote{Phase shift keying.}). The correct decoding
probability $\mathcal{P}_{c}\left(  \mathbf{s}_{8}\right)  $ is the same for
all transmitted signals $\mathbf{s}_{k}$ and it is equal to the volume above
de Voronoi region $V_{8}$ and below to the graphic of $g_{\mathbf{s}_{8}}.$

\begin{center}
\includegraphics{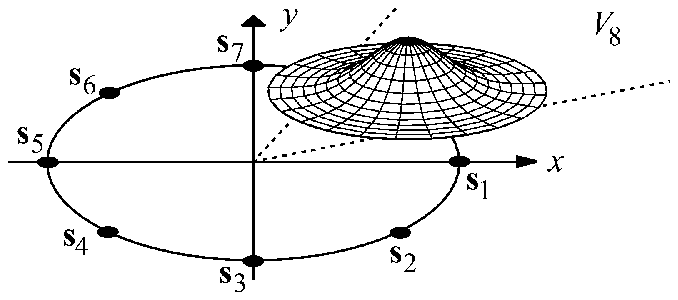}
\end{center}

\begin{center}
\textsc{Figure 1:}\textit{\ Sketch of the graph of a Gaussian pdf associated
to the signal }$\mathbf{s}_{8}$\textit{\ in }$8$\textit{-PSK.}
\end{center}

\bigskip

The next proposition gives an upper bound for (\ref{ProbErroGaussRn}). Before
establishing it, however, it is helpful to introduce the usual notation for
the error function and its complementary counterpart.

The \textit{error function} is defined as:%
\[%
\begin{array}
[c]{cccc}%
\operatorname{erf}: & \mathbb{R}_{+} & \longrightarrow & \left[  0,1\right] \\
& x & \longmapsto & \dfrac{2}{\sqrt{\pi}}%
{\displaystyle\int_{0}^{x}}
e^{-t^{2}}dt
\end{array}
\]
and the \textit{complementary error function} is defined as
$\operatorname{erfc}\left(  x\right)  =1-\operatorname{erf}\left(  x\right)
.$

\begin{prop}
\label{PropLimitSupGaussRn}\cite{Biglieri} Consider a communication system
with a set of $m$ signals in $\mathbb{R}^{n}$ and a transmission channel
affected by AWGN noise. If the $m$ signals are elts, then the error
probability $\mathcal{P}_{e}$ associated to the communication system
satisfies:%
\[
\mathcal{P}_{e}\leq\dfrac{1}{m}%
{\textstyle\sum\limits_{k=1}^{m}}
{\textstyle\sum\limits_{j=1}^{v_{k}}}
\dfrac{1}{2}\operatorname*{erfc}\left(  \dfrac{\left\vert \mathbf{s}%
_{k}-\mathbf{s}_{k_{j}}\right\vert }{2\sqrt{2\sigma^{2}}}\right)
\]
where $\mathbf{s}_{k_{j}},$ $j=1,...,v_{k},$ are the points that determine the
Voronoi region of $\mathbf{s}_{k}.$ \emph{(}\footnote{The signals that
\textquotedblleft determine the Voronoi region\textquotedblright\ of
$\mathbf{s}_{k}$ are the \textquotedblleft neighbors\textquotedblright\ of
$\mathbf{s}_{k},$ that is, the signals $\mathbf{s}_{k_{j}}$ such that the mean
point of the segment $\overline{\mathbf{s}_{k}\mathbf{s}_{k_{j}}}$ is on an
edge of the Voronoi region of $\mathbf{s}_{k}.$}\emph{)}
\end{prop}

The corollary below is helpful when we can not determine which are the signals
that influence the Voronoi region of a signal.

\begin{cor}
In the same conditions of the Proposition \ref{PropLimitSupGaussRn}, we have:%
\[
\mathcal{P}_{e}\leq\dfrac{1}{m}%
{\textstyle\sum\limits_{k=1}^{m}}
{\textstyle\sum\limits_{j\neq k}}
\dfrac{1}{2}\operatorname*{erfc}\left(  \dfrac{\left\vert \mathbf{s}%
_{k}-\mathbf{s}_{j}\right\vert }{2\sqrt{2\sigma^{2}}}\right)  .
\]

\end{cor}

The upper bound given by the next corollary is called \textit{Bhattacharyya
bound.}

\begin{cor}
\emph{(Bhattacharyya Bound)} In the conditions of the Proposition
\ref{PropLimitSupGaussRn}, we have:%
\[
\mathcal{P}_{e}<\dfrac{1}{m}%
{\textstyle\sum\limits_{k=1}^{m}}
{\textstyle\sum\limits_{j\neq k}}
\exp\left(  -\dfrac{\left\vert \mathbf{s}_{k}-\mathbf{s}_{j}\right\vert ^{2}%
}{8\sigma^{2}}\right)  .
\]

\end{cor}

\begin{cor}
\label{CorolGeomUnifGaussRn}In the hypothesis of the Proposition
\ref{PropLimitSupGaussRn}, with the additional condition of the set of signals
being geometrically uniform, we have%
\[
\mathcal{P}_{e}\leq%
{\textstyle\sum\limits_{j=1}^{v_{1}}}
\dfrac{1}{2}\operatorname*{erfc}\left(  \dfrac{\left\vert \mathbf{s}%
_{1}-\mathbf{s}_{1_{j}}\right\vert }{2\sqrt{2\sigma^{2}}}\right)  .
\]

\end{cor}

\section{\label{Sec Gauss Dim 1}Gaussian \textit{pdf} in the One-dimensional
Hyperbolic Space}

The set of positive real numbers $\mathbb{H=R}_{+}^{\ast}$ with the distance%
\[%
\begin{array}
[c]{cccc}%
d_{\mathbb{H}}: & \mathbb{R}_{+}^{\ast}\times\mathbb{R}_{+}^{\ast} &
\longrightarrow & \mathbb{R}_{+}\\
& \left(  x,y\right)  & \longmapsto & \left\vert \ln\dfrac{x}{y}\right\vert
\end{array}
\]
is called the \textit{hyperbolic line}\textbf{\ }or the
\textit{one-dimensional hyperbolic space }and we indicate it by $\mathbb{H}.$

If we consider the set of real numbers $\mathbb{R}$ with the usual Euclidean
distance, $d\left(  x,y\right)  =\left\vert x-y\right\vert ,$ we have that%
\[
T(x)=e^{x}%
\]
is an isometry between $\mathbb{R}$\ and $\mathbb{H}$ with the distance
defined above.

This means that all the inherent properties in $\mathbb{R}$ that depend only
on the usual metric can be transported to $\mathbb{H}$ through the isometry
$T.$ In particular, continuous random variables densities in $\mathbb{R}$ can
be transported to $\mathbb{H},$ because they depend, exclusively, on metric
properties. It is exactly this translation of densities from one context to
the other that we consider below.

\bigskip

Let $X$ be a continuous random variable in $\mathbb{R}$ with Gaussian
\textit{pdf} of mean $\mu$ and variance $\sigma^{2}$:%
\[
g_{\mathbb{R}}\left(  x\right)  =\frac{1}{\sqrt{2\pi\sigma^{2}}}\exp\left(
-\frac{d^{2}\left(  x,\mu\right)  }{2\sigma^{2}}\right)  .
\]

We have, naturally,
\[
\int_{-\infty}^{+\infty}g_{\mathbb{R}}\left(  x\right)  dx=1.
\]

To find the \textit{pdf} $g_{\mathbb{H}},$ in $\mathbb{H},$ which corresponds
to the Gaussian \textit{pdf }$g_{\mathbb{R}}$ we use the differentiable
isometry $T$ and deduce the hyperbolic arc length element:%
\[
d_{\mathbb{H}}x=\left\vert \dfrac{dT^{-1}}{dx}\left(  x\right)  \right\vert
dx=\dfrac{1}{x}dx,
\]
and then:%
\[
g_{\mathbb{H}}\left(  x\right)  =\frac{1}{\sqrt{2\pi\sigma^{2}}}\exp\left(
-\frac{d_{\mathbb{H}}^{2}\left(  x,e^{\mu}\right)  }{2\sigma^{2}}\right)  ,
\]
for the one univariated hyperbolic Gaussian \textit{pdf}, which satisfies:%
\[
\int_{0}^{+\infty}g_{\mathbb{H}}\left(  x\right)  d_{\mathbb{H}}x=\int
_{0}^{+\infty}\dfrac{g_{\mathbb{H}}\left(  x\right)  }{x}dx=\int_{-\infty
}^{+\infty}g_{\mathbb{R}}\left(  x\right)  dx=1.
\]

We remark that the hyperbolic Gaussian \textit{pdf}, $g_{\mathbb{H}},$ when
described in Euclidean terms, is given by $p\left(  x\right)  =\dfrac
{g_{\mathbb{H}}\left(  x\right)  }{x},$ that is, the usual lognormal
\textit{pdf}.

\bigskip

We define the \textit{symmetrical mean} of $g_{\mathbb{H}}$ by $\mu
_{\mathbb{H}}=e^{\mu},$ where $\mu$ is the mean of the Gaussian \textit{pdf}
in $\mathbb{R}$ and the \textit{variance} of $g_{\mathbb{H}}$ by:%
\[
\sigma_{\mathbb{H}}^{2}=\int_{\mathbb{H}}d_{\mathbb{H}}^{2}\left(
x,\mu_{\mathbb{H}}\right)  g_{\mathbb{H}}\left(  x\right)  d_{\mathbb{H}%
}x=\int_{\mathbb{H}}\left(  \ln\frac{x}{\mu_{\mathbb{H}}}\right)
^{2}g_{\mathbb{H}}\left(  x\right)  d_{\mathbb{H}}x,
\]
in analogy with the Euclidean Gaussian variance. And hence, for the hyperbolic
\textit{pdf} $g_{\mathbb{H}}$ with mean $\mu_{\mathbb{H}}=e^{\mu}$ defined
above, a straightforward calculation shows that $\sigma_{\mathbb{H}}%
^{2}=\sigma^{2}.$ Besides, we get the same results that hold for the Gaussian
\textit{pdf} in $\mathbb{R}$:

\begin{itemize}
\item $\mu_{\mathbb{H}}$ is the symmetrical point of the division of the graph
of $g_{\mathbb{H}}$ concerning \ hyperbolic distances. (i.e. $g_{\mathbb{H}}$
assume the same values at points at the same hyperbolic distance on the left
and on the right of $\mu_{\mathbb{H}}.$):%
\[
\int_{0}^{\mu_{\mathbb{H}}}g_{\mathbb{H}}\left(  x\right)  d_{\mathbb{H}%
}x=\int_{\mu_{\mathbb{H}}}^{+\infty}g_{\mathbb{H}}\left(  x\right)
d_{\mathbb{H}}x=\dfrac{1}{2}%
\]
and%
\[%
{\displaystyle\int_{0}^{e^{\mu-z}}}
g_{\mathbb{H}}\left(  x\right)  d_{\mathbb{H}}x=%
{\displaystyle\int_{e^{\mu+z}}^{+\infty}}
g_{\mathbb{H}}\left(  x\right)  d_{\mathbb{H}}x,\text{ }\forall z\in
\mathbb{R}.
\]

\item The maximum value of the \textit{pdf} occurs at the mean and we also
have:%
\[
\int_{\mu_{\mathbb{H}}e^{-\sigma}}^{\mu_{\mathbb{H}}}g_{\mathbb{H}}\left(
x\right)  d_{\mathbb{H}}x=\int_{\mu_{\mathbb{H}}}^{\mu_{\mathbb{H}}e^{\sigma}%
}g_{\mathbb{H}}\left(  x\right)  d_{\mathbb{H}}x=\frac{1}{2}\operatorname{erf}%
\left(  \frac{1}{2}\sqrt{2}\right)  =\int_{\mu-\sigma}^{\mu+\sigma
}g_{\mathbb{R}}\left(  x\right)  dx,
\]
what will imply that \textit{independent} of the value of $\sigma_{\mathbb{H}%
},$ we have approximately $68,27\%$ of the distribution $g_{\mathbb{H}}$
between the points $\mu_{\mathbb{H}}e^{-\sigma}$ and $\mu_{\mathbb{H}%
}e^{\sigma},$ which are (hyperbolic) equidistant from $\mu_{\mathbb{H}}.$
\end{itemize}

\bigskip

As we have pointed out, the hyperbolic Gaussian \textit{pdf,} when described
in Euclidean terms, is precisely the lognormal \textit{pdf.} So, under the
hyperbolic view, a lognormal \textit{pdf} have symmetry with respect to equal
(hyperbolic) distances from the mean.

Figure 2 shows the \textit{Euclidean graphs}\footnote{We understand by the
\textit{Euclidean graph} of a \textit{pdf} $g$ with domain in $\mathbb{H}^{n}$
or $\mathbb{B}^{n}$ as being the set of points $\left(  \mathbf{x},g\left(
\mathbf{x}\right)  \right)  $ of the space $\mathbb{R}^{n+1}$ such that
$\mathbf{x}$ is in domain of $g.$} of the two Gaussian \textit{pdf's}
$g_{\mathbb{H}}$ of variance $\sigma^{2}=0,1;$ one with mean $\mu_{\mathbb{H}%
}=e^{0}=1$ and another with mean $\mu_{\mathbb{H}}=e^{-1}.$

\begin{center}
\includegraphics{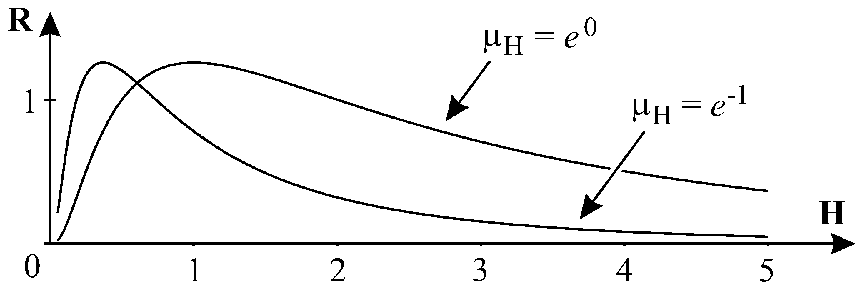}
\end{center}

\begin{center}
\textsc{Figure 2:}\textit{\ Euclidean graphs of the hyperbolic Gaussian
densities of variance }$\sigma^{2}=0,1$ \textit{in }$\mathbb{H}.$
\end{center}

\bigskip

In the study of one-dimensional hyperbolic Gaussian \textit{pdf} done here
there is nothing really new:\textit{\ }it is just the lognormal \textit{pdf}
that can be viewed as symmetric in terms of the hyperbolic distance. But the
discussion presented here paves the way for considering \textit{pdf }in
dimensions greater or equal to two. Since in those dimensions there is no
isometry between Euclidean an hyperbolic spaces we have no equivalence
whatsoever and the hyperbolic \textit{pdf} must then be dealt strictly in
hyperbolic terms.

\section{\label{Sec Gauss Lim 2}Gaussian \textit{pdf} in Hyperbolic Spaces of
Dimension Greater or Equal than Two}

We consider here the two Poincar\'{e}'s Euclidean models for hyperbolic
$n$-dimensional geometry $\left(  n\geq2\right)  $:%
\[
\mathbb{M}_{H}^{n}=\left\{  \mathbf{x}=\left(  x_{1},...,x_{n}\right)
:x_{i}\in\mathbb{R},\text{ }x_{n}>0\right\}  ,
\]
called upper half-space model, and:
\[
\mathbb{M}_{B}^{n}=\left\{  \mathbf{x}=\left(  x_{1},...,x_{n}\right)
:x_{i}\in\mathbb{R},\text{ }x_{1}^{2}+...+x_{n}^{2}<1\right\}  ,
\]
called unitary ball model, associated with the Riemannian metrics:%
\[
ds_{H}=\dfrac{\sqrt{dx_{1}^{2}+...+dx_{n}^{2}}}{x_{n}}\text{ and }%
ds_{B}=\dfrac{2\sqrt{dx_{1}^{2}+...+dx_{n}^{2}}}{1-\left(  x_{1}^{2}%
+...+x_{n}^{2}\right)  },
\]
respectively.

The distance between two points in those models \cite{Beardon} will be then
given by:%
\[
d_{\mathbb{M}_{H}^{n}}\left(  \mathbf{x},\mathbf{y}\right)  =\ln
\dfrac{\left\vert \mathbf{x}-\overline{\mathbf{y}}\right\vert +\left\vert
\mathbf{x}-\mathbf{y}\right\vert }{\left\vert \mathbf{x}-\overline{\mathbf{y}%
}\right\vert -\left\vert \mathbf{x}-\mathbf{y}\right\vert }%
\]
and%
\[
d_{\mathbb{M}_{B}^{n}}\left(  \mathbf{x},\mathbf{y}\right)  =\ln\dfrac
{\sqrt{\left(  1-\left\vert \mathbf{x}\right\vert ^{2}\right)  \left(
1-\left\vert \mathbf{y}\right\vert ^{2}\right)  +\left\vert \mathbf{x}%
-\mathbf{y}\right\vert ^{2}}+\left\vert \mathbf{x}-\mathbf{y}\right\vert
}{\sqrt{\left(  1-\left\vert \mathbf{x}\right\vert ^{2}\right)  \left(
1-\left\vert \mathbf{y}\right\vert ^{2}\right)  +\left\vert \mathbf{x}%
-\mathbf{y}\right\vert ^{2}}-\left\vert \mathbf{x}-\mathbf{y}\right\vert }.
\]

In two dimensional hyperbolic space, $n=2,$ we can simplify the last
expression:%
\[
d_{\mathbb{M}_{B}^{2}}\left(  \mathbf{x},\mathbf{y}\right)  =\ln
\dfrac{\left\vert 1-\mathbf{x}\overline{\mathbf{y}}\right\vert +\left\vert
\mathbf{x}-\mathbf{y}\right\vert }{\left\vert 1-\mathbf{x}\overline
{\mathbf{y}}\right\vert -\left\vert \mathbf{x}-\mathbf{y}\right\vert },
\]
where $\overline{\mathbf{y}}$ is the complex conjugate of $\mathbf{y}.$

These two models are isometric and hence everything that is done in one can be
translated for other in terms of the isometries:%
\[%
\begin{array}
[t]{cccc}%
I: & \mathbb{M}_{H}^{n} & \longrightarrow & \mathbb{M}_{B}^{n}\\
& \mathbf{x} & \longmapsto & 2\dfrac{\overline{\mathbf{x}+\mathbf{p}}%
}{\left\vert \mathbf{x}+\mathbf{p}\right\vert ^{2}}+\mathbf{p}%
\end{array}
\text{ and }%
\begin{array}
[t]{cccc}%
I^{-1}: & \mathbb{M}_{B}^{n} & \longrightarrow & \mathbb{M}_{H}^{n}\\
& \mathbf{x} & \longmapsto & 2\dfrac{\overline{\mathbf{x}-\mathbf{p}}%
}{\left\vert \mathbf{x}-\mathbf{p}\right\vert ^{2}}-\mathbf{p}%
\end{array}
,
\]
where $\mathbf{p}=\left(  0,...,0,1\right)  \in\mathbb{R}^{n}.$

In order to establish a \textit{pdf} in $n$-dimensional hyperbolic spaces, we
call $\mathbb{M}_{H}^{n}$ or $\mathbb{M}_{B}^{n}$ of $\mathbb{H}^{n}.$

For dimension $n,$ $n\geq2,$ the striking point is the non-existence of an
isometry between $\mathbb{R}^{n}$ and the $n$-dimensional hyperbolic space
$\mathbb{H}^{n}$. Therefore, there is no natural way to define, by means of
isometries, a \textit{pdf} in $\mathbb{H}^{n}$ \textquotedblleft
equivalent\textquotedblright\ to the \textit{Gaussian pdf} of $\mathbb{R}^{n}
$:%
\[
g_{\mathbb{R}^{n}}\left(  \mathbf{x}\right)  =\dfrac{1}{\sqrt{\left(
2\pi\sigma^{2}\right)  ^{n}}}\exp\left(  -\dfrac{d^{2}\left(  \mathbf{x}%
,\mathbf{\mu}\right)  }{2\sigma^{2}}\right)  ,
\]
where $\mathbf{x}=\left(  x_{1},...,x_{n}\right)  \in\mathbb{R}^{n},$
$\mathbf{\mu}=\left(  \mu_{1},...,\mu_{n}\right)  $ are the mean and
$\sigma^{2}$ the variance of the \textit{pdf} $g_{\mathbb{R}^{n}}.$

However, it is possible to define a \textit{pdf} for the hyperbolic space that
possesses the same geometric characteristics of the Gaussian \textit{pdf} in
the Euclidean space. The first fact to be observed in $g_{\mathbb{R}^{n}}$ is
that the dimension does not influence in the factor $\frac{1}{2\sigma^{2}}$
that multiplies $-d^{2}$ and that $\mu$ is the point of radial symmetry of
$g_{\mathbb{R}^{n}}.$ The deduction of the proper definition for
$g_{\mathbb{H}}$ that we developed in the previous section shows us that these
properties should be preserved in $g_{\mathbb{H}^{n}},$ as well as the
equality between the Euclidean and the associated hyperbolic variance;
$\sigma_{\mathbb{H}^{n}}^{2}=\sigma^{2}.$ Those geometric facts are the
support for the following definition of $n$-dimensional Gaussian \textit{pdf}
in $n$-dimensional hyperbolic space. Thus, we define:%
\[
g_{\mathbb{H}^{n}}\left(  \mathbf{x}\right)  =k_{\mathbb{H}^{n}}\exp\left(
-\dfrac{1}{2\sigma^{2}}d_{\mathbb{H}^{n}}^{2}\left(  \mathbf{x},\mathbf{\mu
}\right)  \right)  ,
\]
where $k_{\mathbb{H}^{n}}$ is the normalization factor, i.e., a constant to be
determined in order to $%
{\displaystyle\int\nolimits_{\mathbb{H}^{n}}}
g_{\mathbb{H}^{n}}\left(  \mathbf{x}\right)  dV_{\mathbb{H}^{n}}=1,$
$dV_{\mathbb{H}^{n}}$ is the volume element in $\mathbb{H}^{n},$ $\mathbf{\mu
}$ is the mean and $\sigma^{2}$ is the variance of the \textit{pdf}
$g_{\mathbb{H}^{n}}.$

In the development to find $k_{\mathbb{H}^{n}}$ that we do next, we opted for
the unitary ball model due to its simplicity in terms of variable change in
this case. Its important to point out that $k_{\mathbb{H}^{n}}$ is independent
of the used model.

Let us consider, without loss of generality, the hyperbolic mean as
$\mathbf{\mu}=\mathbf{0}$. Therefore, we must have:\scalefont{0.8}
\begin{align*}
\int\nolimits_{\mathbb{H}^{n}}k_{\mathbb{H}^{n}}\exp\left(  -\dfrac
{d_{\mathbb{H}^{n}}^{2}\left(  \mathbf{x},\mathbf{0}\right)  }{2\sigma^{2}%
}\right)  dV_{\mathbb{H}^{n}}  &  =1\Rightarrow\\
\int\nolimits_{B^{n}}k_{\mathbb{H}^{n}}\exp\left(  -\dfrac{d_{\mathbb{H}^{n}%
}^{2}\left(  \mathbf{x},\mathbf{0}\right)  }{2\sigma^{2}}\right)  \left(
\dfrac{2}{1-\left\vert \mathbf{x}\right\vert ^{2}}\right)  ^{n}dV_{\mathbb{R}%
^{n}}  &  =1\Rightarrow\\
\int\nolimits_{-1}^{1}\int\nolimits_{-\sqrt{1-x_{1}^{2}}}^{\sqrt{1-x_{1}^{2}}%
}...\int\nolimits_{-\sqrt{1-x_{1}^{2}-...-x_{n}^{2}}}^{\sqrt{1-x_{1}%
^{2}-...-x_{n}^{2}}}k_{\mathbb{H}^{n}}e^{-\frac{1}{2\sigma^{2}}\ln^{2}%
\frac{1+\sqrt{x_{1}^{2}+...+x_{n}^{2}}}{1-\sqrt{x_{1}^{2}+...+x_{n}^{2}}}%
}\left(  \dfrac{2}{1-x_{1}^{2}-...-x_{n}^{2}}\right)  ^{n}dx_{1}...dx_{n}  &
=1.
\end{align*}
\scalefont{1.25}

We introduce the following hyperspherical coordinates to simplify the
development:%
\begin{align*}
x_{n}  &  =r\cos\alpha_{1}\\
x_{n-1}  &  =r\sin\alpha_{1}\cos\alpha_{2}\\
x_{n-2}  &  =r\sin\alpha_{1}\sin\alpha_{2}\cos\alpha_{3}\\
x_{n-3}  &  =r\sin\alpha_{1}\sin\alpha_{2}\sin\alpha_{3}\cos\alpha_{4}\\
&  \vdots\\
x_{3}  &  =x_{n-\left(  n-3\right)  }=r\sin\alpha_{1}\sin\alpha_{2}\sin
\alpha_{3}...\sin\alpha_{n-3}\cos\alpha_{n-2}\\
x_{2}  &  =x_{n-\left(  n-2\right)  }=r\sin\alpha_{1}\sin\alpha_{2}\sin
\alpha_{3}...\sin\alpha_{n-3}\sin\alpha_{n-2}\cos\alpha_{n-1}\\
x_{1}  &  =x_{n-\left(  n-1\right)  }=r\sin\alpha_{1}\sin\alpha_{2}\sin
\alpha_{3}...\sin\alpha_{n-3}\sin\alpha_{n-2}\sin\alpha_{n-1}%
\end{align*}
where: $r\in\mathbb{R}_{+},$ $0\leq\alpha_{n-1}<2\pi,$ and $0\leq\alpha
_{1},...,\alpha_{n-2}\leq\pi.$

The Jacobian of the coordinate change is given by:%
\[
\dfrac{\partial\left(  x_{1},...,x_{n}\right)  }{\partial\left(  r,\alpha
_{1},...,\alpha_{n-1}\right)  }=r^{n-1}\sin^{n-2}\alpha_{1}\sin^{n-3}%
\alpha_{2}...\sin^{2}\alpha_{n-3}\sin\alpha_{n-2}.
\]

Thus, taking:%
\[
drd\alpha_{1}...d\alpha_{n-2}d\alpha_{n-1}=dV,
\]
the above integral can be written as:\scalefont{0.8}
\begin{align*}
\int\nolimits_{0}^{2\pi}\int\nolimits_{0}^{\pi}...\int\nolimits_{0}^{\pi}%
\int\nolimits_{0}^{1}k_{\mathbb{B}^{n}}\exp\left(  -\dfrac{1}{2\sigma^{2}}%
\ln^{2}\dfrac{1+r}{1-r}\right)  \left(  \dfrac{2}{1-r^{2}}\right)
^{n}\left\vert \dfrac{\partial\left(  x_{1},...,x_{n}\right)  }{\partial
\left(  r,\alpha_{1},...,\alpha_{n-1}\right)  }\right\vert dV  &
=1\Rightarrow\\
\int\nolimits_{0}^{2\pi}\int\nolimits_{0}^{\pi}...\int\nolimits_{0}^{\pi
}\left(  \int\nolimits_{0}^{1}k_{\mathbb{H}^{n}}\exp\left(  -\dfrac{1}%
{2\sigma^{2}}\ln^{2}\dfrac{1+r}{1-r}\right)  \left(  \dfrac{2}{1-r^{2}%
}\right)  ^{n}r^{n-1}dr\right)  \dfrac{\left\vert \frac{\partial\left(
x_{1},...,x_{n}\right)  }{\partial\left(  r,\alpha_{1},...,\alpha
_{n-1}\right)  }\right\vert }{r^{n-1}}\dfrac{dV}{dr}  &  =1\Rightarrow\\
\left(  \int\nolimits_{0}^{1}k_{\mathbb{H}^{n}}\exp\left(  -\dfrac{1}%
{2\sigma^{2}}\ln^{2}\dfrac{1+r}{1-r}\right)  \left(  \dfrac{2}{1-r^{2}%
}\right)  ^{n}r^{n-1}dr\right)  \int\nolimits_{0}^{2\pi}\int\nolimits_{0}%
^{\pi}...\int\nolimits_{0}^{\pi}\dfrac{\left\vert \frac{\partial\left(
x_{1},...,x_{n}\right)  }{\partial\left(  r,\alpha_{1},...,\alpha
_{n-1}\right)  }\right\vert }{r^{n-1}}\dfrac{dV}{dr}  &  =1\Rightarrow\\
\left(  \int\nolimits_{0}^{1}k_{\mathbb{H}^{n}}e^{-\frac{1}{2\sigma^{2}}%
\ln^{2}\frac{1+r}{1-r}}\left(  \tfrac{2}{1-r^{2}}\right)  ^{n}r^{n-1}%
dr\right)  n\int\nolimits_{0}^{2\pi}\int\nolimits_{0}^{\pi}...\int
\nolimits_{0}^{\pi}\left(  \int\nolimits_{0}^{1}r^{n-1}dr\right)
\tfrac{\left\vert \frac{\partial\left(  x_{1},...,x_{n}\right)  }%
{\partial\left(  r,\alpha_{1},...,\alpha_{n-1}\right)  }\right\vert }{r^{n-1}%
}\dfrac{dV}{dr}  &  =1\Rightarrow\\
\left(  \int\nolimits_{0}^{1}k_{\mathbb{H}^{n}}e^{-\frac{1}{2\sigma^{2}}%
\ln^{2}\frac{1+r}{1-r}}\left(  \dfrac{2}{1-r^{2}}\right)  ^{n}r^{n-1}%
dr\right)  n\int\nolimits_{0}^{2\pi}\int\nolimits_{0}^{\pi}...\int
\nolimits_{0}^{\pi}\int\nolimits_{0}^{1}\left\vert \dfrac{\partial\left(
x_{1},...,x_{n}\right)  }{\partial\left(  r,\alpha_{1},...,\alpha
_{n-1}\right)  }\right\vert dV  &  =1.
\end{align*}
\scalefont{1.25}

But:%
\[
\int\nolimits_{0}^{2\pi}\int\nolimits_{0}^{\pi}...\int\nolimits_{0}^{\pi}%
\int\nolimits_{0}^{1}\left\vert \dfrac{\partial\left(  x_{1},...,x_{n}\right)
}{\partial\left(  r,\alpha_{1},...,\alpha_{n-1}\right)  }\right\vert dV
\]
is exactly the volume $V_{B_{1}^{n}}$ of the hyperball $B_{1}^{n}$ of
dimension $n$ and radius $1.$

Therefore,
\[
\int\nolimits_{0}^{1}k_{\mathbb{H}^{n}}\exp\left(  -\dfrac{1}{2\sigma^{2}}%
\ln^{2}\dfrac{1+r}{1-r}\right)  \left(  \dfrac{2}{1-r^{2}}\right)  ^{n}%
r^{n-1}dr=\dfrac{1}{nV_{B_{1}^{n}}}.
\]

We will need the next proposition, which can be found, for example, in
\cite{Sloane}.

\begin{prop}
\emph{(Volume of a ball in the }$n$\emph{-dimensional Euclidean space
\cite{Sloane}) }Let $B_{R}^{n}$ be the ball of dimension $n$ and radius $R$ in
the Euclidean space. Then, the volume of $B_{R}^{n},$ denoted by $V_{B_{R}%
^{n}},$ is given by%
\begin{align*}
V_{B_{R}^{n}}  &  =\dfrac{R^{n}\pi^{\frac{n}{2}}}{\left(  \frac{n}{2}\right)
!}\text{ if }n\text{ if even}\\
V_{B_{R}^{n}}  &  =\dfrac{R^{n}\pi^{\frac{n-1}{2}}2^{n}\left(  \frac{n-1}%
{2}\right)  !}{n!}\text{ if }n\text{ is odd.}%
\end{align*}

\end{prop}

Coming back to the hyperbolic \textit{pdf}, let us consider another change of
coordinates:%
\[
u=\ln\dfrac{1+r}{1-r}\Longrightarrow\dfrac{du}{dr}=\dfrac{2}{1-r^{2}}\text{
and }r=\dfrac{e^{u}-1}{e^{u}+1}.
\]

Thus,
\begin{gather*}%
{\displaystyle\int_{0}^{+\infty}}
k_{\mathbb{H}^{n}}\exp\left(  -\frac{u^{2}}{2\sigma^{2}}\right)  \left(
\dfrac{e^{2u}-1}{2e^{u}}\right)  ^{n-1}du=\dfrac{1}{nV_{B_{1}^{n}}}%
\Rightarrow\\
k_{\mathbb{H}^{n}}=\dfrac{1}{nV_{B_{1}^{n}}%
{\displaystyle\int_{0}^{+\infty}}
\exp\left(  -\frac{u^{2}}{2\sigma^{2}}\right)  \left(  \frac{e^{2u}-1}{2e^{u}%
}\right)  ^{n-1}du},
\end{gather*}
that is:%
\[
k_{\mathbb{H}^{n}}=\dfrac{1}{nV_{B_{1}^{n}}%
{\displaystyle\int_{0}^{+\infty}}
\exp\left(  -\frac{u^{2}}{2\sigma^{2}}\right)  \left(  \sinh u\right)
^{n-1}du}.
\]

We remark that, like in the Euclidean environment, $k_{\mathbb{H}^{n}}$
depends on the variance $\sigma^{2}$ but does not depend on the mean
$\mathbf{\mu}.$

Having obtained the \textit{pdf} for the unitary ball model we can, if
convenient, transport it to the half-space model by means of the isometry $I,
$ that is, the \textit{pdf} $g_{1}$ in the ball must be \textquotedblleft
isometric\textquotedblright\ to the \textit{pdf} $g_{2}$ in the half-space:%
\[
g_{2}\left(  \mathbf{y}\right)  =g_{1}\left(  I\left(  \mathbf{y}\right)
\right)
\]

With these considerations, we can give a formal definition for the Gaussian
\textit{pdf} in the hyperbolic space $\mathbb{H}^{n}.$

\begin{defn}
Let $\mathbb{H}^{n}$ be a Poincar\'{e}'s model for the hyperbolic geometry. We
define the \emph{\ }$n$\emph{-dimensional hyperbolic Gaussian density} with
mean $\mathbf{\mu}$ and variance $\sigma^{2}$ in $\mathbb{H}^{n}$ as being%
\[
g_{\mathbb{H}^{n}}\left(  \mathbf{x}\right)  =k_{\mathbb{H}^{n}}\exp\left(
-\frac{d_{\mathbb{H}^{n}}^{2}\left(  \mathbf{x},\mathbf{\mu}\right)  }%
{2\sigma^{2}}\right)
\]
with%
\[
k_{\mathbb{H}^{n}}=\frac{1}{nV_{B_{1}^{n}}%
{\displaystyle\int_{0}^{+\infty}}
\exp\left(  -\frac{u^{2}}{2\sigma^{2}}\right)  \left(  \sinh u\right)
^{n-1}du}.
\]

\end{defn}

For $n=2$ we deduce an analytic expression for $k_{\mathbb{H}^{2}}$ as stated
in the next proposition. For $n>2$ this constant can be obtained through
numeric processes with very good accuracy.

\begin{prop}
Let $g_{\mathbb{H}^{2}}\left(  \mathbf{x}\right)  =k_{\mathbb{H}^{2}}%
\exp\left(  -\frac{d_{\mathbb{H}^{2}}^{2}\left(  \mathbf{x},\mathbf{\mu
}\right)  }{2\sigma^{2}}\right)  $ be the two-dimensional hyperbolic Gaussian
pdf in $\mathbb{H}^{2}.$ Then $k_{\mathbb{H}^{2}}=\frac{1}{\sqrt{2\sigma
^{2}\pi^{3}}\exp\left(  \frac{\sigma^{2}}{2}\right)  \operatorname{erf}\left(
\sqrt{\frac{\sigma^{2}}{2}}\right)  }.$
\end{prop}

Three important remarks remain, which extend the previous discussion of the
$1$-dimensional case:

\begin{itemize}
\item the mean $\mathbf{\mu}$ is the maximum in the Euclidean graph of
$g_{\mathbb{H}^{n}},$ that is, $k_{\mathbb{H}^{n}}$ is the maximum value of
$g_{\mathbb{H}^{n}}.$

\item the hyperbolic Gaussian \textit{pdf} is radially symmetric, that is, the
level hypersurfaces\footnote{If $n=2$: level curves. If $n=3$: level
surfaces.} of the Euclidean graphic of $g_{\mathbb{H}^{n}}$ are exactly
hyperbolic hyperspheres\footnote{If $n=2$: hyperbolic circumferences. If
$n=3$: hyperbolic spheres.} with the center at the mean. In fact:

Let $c\in\mathbb{R}_{+}^{\ast}$ such that%
\[
g_{\mathbb{H}^{n}}\left(  \mathbf{y}\right)  =c.
\]
Thus, $c\leq k_{\mathbb{H}^{n}}$ due to the above remark and, therefore,%
\[
-2\sigma^{2}\ln\dfrac{c}{k_{\mathbb{H}^{n}}}\geq0.
\]
From $g_{\mathbb{H}^{n}}\left(  \mathbf{y}\right)  =c$ we have%
\[
d_{\mathbb{H}^{n}}\left(  \mathbf{y,\mu}\right)  =\sqrt{-2\sigma^{2}\ln
\dfrac{c}{k_{\mathbb{H}^{n}}}},
\]
that is, $g_{\mathbb{H}^{n}}\left(  \mathbf{y}\right)  =c$ is a hyperbolic
hypersphere of center $\mathbf{\mu}$ and radius $\sqrt{-2\sigma^{2}\ln\frac
{c}{k_{\mathbb{H}^{n}}}}.$
\end{itemize}

Figure 3 shows the Euclidean graphics of $g_{\mathbb{H}^{2}}$ for both models
(moved upward) and some level curves.

\begin{center}
\includegraphics{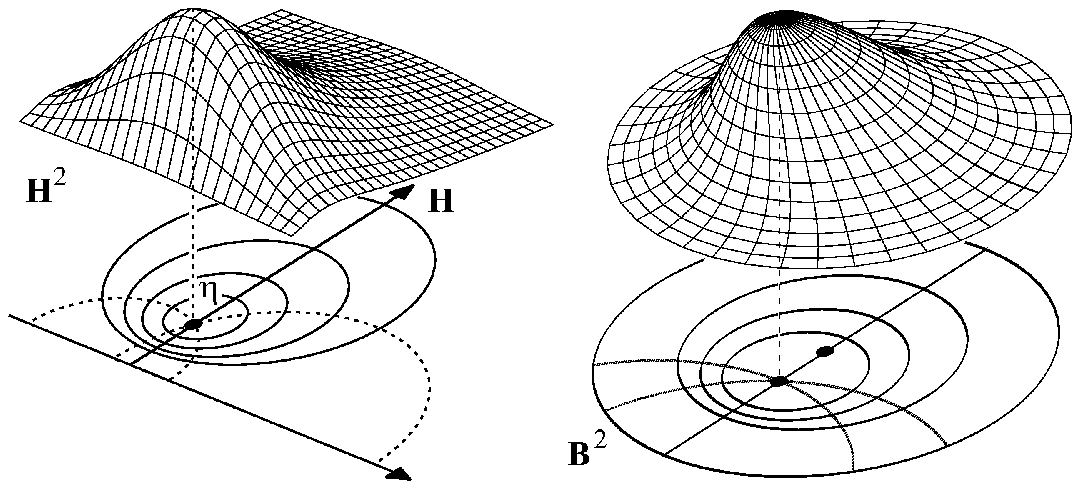}
\end{center}

\begin{center}
\textsc{Figure 3:}\textit{\ Euclidean graphs of }$g_{\mathbb{H}^{2}}%
$\textit{\ (moved upward) with means }$\mathbf{\mu}$\textit{\ (ball model) and
}$\mathbf{\eta}$\textit{\ (half-space model). The hyperbolic circumferences
are the level curves.}
\end{center}

\bigskip

For Euclidean Gaussian \textit{pdf's} with a fixed variance, we have
$k_{\mathbb{R}^{n}}=\left(  k_{\mathbb{R}}\right)  ^{n},$ since a
$n$-dimensional \textit{pdf} can be written as a product of one-dimensional
\textit{pdf's}, what is helpful in estimating error probability. The analogous
equality does not happen in the hyperbolic space (due to the lack of a
vectorial structure in the $n$-dimensional hyperbolic space which is
compactible with the hyperbolic metric), and so the establishment of bounds
for error probability will be even more necessary.

\section{\label{Sec Hyp Limit}An Upper Bound for the Hyperbolic Error
Probability}

As it is well know, estimating error probability in higher dimensional spaces
is usually a hard task even in the Euclidean case. Since the \textit{pdf's}
expressions in hyperbolic space are more complicated it is very important to
set good upper bounds for it. We have emphasized that the hyperbolic Gaussian
\textit{pdf} is not a product of \textit{pfd's} of smaller dimensions, which,
certainly, complicates the search for a bound. However, based on the fact that
in a modulation system with $m$ signals the conditional error probability
$\mathcal{P}_{e}\left(  \mathbf{s}_{k}\right)  $ associated to the
transmission of a signal $\mathbf{s}_{j}$ always satisfies the condition%
\[
\mathcal{P}_{e}\left(  \mathbf{s}_{k}\right)  \leq%
{\textstyle\sum\limits_{m\neq k}}
P\left(  \mathbf{s}_{k},\mathbf{s}_{m}\right)  ,
\]
where $P\left(  \mathbf{s}_{k},\mathbf{s}_{m}\right)  $ is the error
probability $P\left(  \mathbf{s}_{k},\mathbf{s}_{m}\right)  =P\left(
\mathbf{s}_{k}\right)  P\left(  \mathbf{s}_{m}|\mathbf{s}_{k}\right)  ,$ we
note that this upper bound for $\mathcal{P}_{e}\left(  \mathbf{s}_{k}\right)
$ is given by the sum of parts, each one depending just a pair of signals.

Each pair of signals determines a geodesic (straight line with the considered
metric) in $\mathbb{H}^{n}.$ But any geodesic in $\mathbb{H}^{n}$ is isometric
to $\mathbb{H},$ the one-dimensional hyperbolic space. Therefore, we will
suppose that each pair of signals is submitted to a Gaussian one-dimensional
noise in an independent way. This noise, as in Euclidean case, will be
considered as modeled by a Gaussian \textit{pdf} that, in this case, is the
lognormal \textit{pdf} (Section \ref{Sec Gauss Dim 1}).

Our aim now is to obtain upper bounds for the error probability in hyperbolic
case comparable to that developed in the Euclidean case (Proposition
\ref{PropLimitSupGaussRn}).

We begin with the following technical lemma.

\begin{lem}
\label{LemaErfHiperbolico} Let $a\in\mathbb{R}_{+},$ $b\in\mathbb{R}$ and
$c\in\mathbb{R}_{+}^{\ast}.$ Then
\[
\int_{a}^{+\infty}y^{b-c\ln y}dy=\sqrt{\dfrac{\pi}{4c}}\exp\left(
\frac{\left(  b+1\right)  ^{2}}{4c}\right)  \operatorname*{erfc}\left(
\dfrac{2c\ln a-b-1}{\sqrt{4c}}\right)  .
\]

\end{lem}

\begin{thm}
\label{TeoremaEstimador}Let a set of signals $\mathcal{S}=\left\{
\mathbf{s}_{1},...,\mathbf{s}_{m}\right\}  $ in $\mathbb{H}^{n},$ $n\geq2,$
used in a communication system with channel affected by hyperbolic Gaussian
noise. If $\mathcal{S}$ have \textit{equally likely to be transmitted signals}
\emph{(elts)}, then the error probability associated to $\mathcal{S}$
satisfies%
\[
\mathcal{P}_{e}\leq\dfrac{1}{m}%
{\textstyle\sum\limits_{k=1}^{m}}
{\textstyle\sum\limits_{j=1}^{v_{k}}}
\dfrac{1}{2}\operatorname*{erfc}\left(  \frac{d_{\mathbb{H}^{n}}\left(
\mathbf{s}_{k},\mathbf{s}_{k_{j}}\right)  }{2\sqrt{2\sigma^{2}}}\right)  ,
\]
where $\sigma^{2}$ is the noise variance and $v_{k}$ is the number of signals
$\mathbf{s}_{k_{j}}$ that determines a face boundary\footnote{See the footnote
of the Proposition \ref{PropLimitSupGaussRn}.} in the Voronoi region of
$\mathbf{s}_{k}.$
\end{thm}

\noindent\textbf{Proof}

We choose the half-space model for this proof.

Due to $\mathcal{S}$ have \textit{elts}, we can fix an index, for example
$k=1,$ and deduce an upper bound for the error probability $P\left(
\mathbf{s}_{1},\mathbf{s}_{p}\right)  $ of receiving $\mathbf{s}_{p}$ when
$\mathbf{s}_{1}$ is transmitted, $p\neq1.$ However, not all the $p$'$s$ are
necessary for the calculation of the bound for $\mathcal{P}_{e}.$ We consider
some restrictions.

Let us take the Voronoi region $V_{1}$ of $\mathbf{s}_{1}.$ As the amount of
signals in the constellation $\mathcal{S}$ is finite, then $V_{1}$ is a
hyperbolic polytope ($n$-dimensional polyhedron not necessarily compact)
defined by the intersection of a finite number of $\left(  n-1\right)
$-dimensional hyperplanes. Each one of these hyperplanes is a
\textquotedblleft bisector\textquotedblright\ between $\mathbf{s}_{1}$ and
$\mathbf{s}_{j}$ for some $j$ and some of them will determine a face on
$V_{1}.$

As a simple example, the signal $\mathbf{s}_{1}$ in the Euclidean plane set
called $4$\textit{-PSK} ($\mathbf{s}_{1}=\left(  1,0\right)  ,$ $\mathbf{s}%
_{2}=\left(  0,-1\right)  ,$ $\mathbf{s}_{3}=\left(  -1,0\right)  ,$
$\mathbf{s}_{4}\left(  0,1\right)  $) has a Voronoi region $V_{1}$ defined by
the bisectors of $\overline{\mathbf{s}_{1}\mathbf{s}_{2}}$ and $\overline
{\mathbf{s}_{1}\mathbf{s}_{4}},$ but the bisector of $\overline{\mathbf{s}%
_{1}\mathbf{s}_{3}}$ does not intercept the interior of $V_{1}$ and,
therefore, it can be discarded. Therefore, we only consider the points whose
bisectors determine side in $V_{1}$ and, without loss of generality, let us
suppose that these are $\mathbf{s}_{1_{1}},...,\mathbf{s}_{1_{v_{1}}},$
$v_{1}\leq M. $

The signals $\mathbf{s}_{1}$ and $\mathbf{s}_{1_{j}}$ are \textit{elts}, so
they are submitted to the unidimensional Gaussian noise of the same variance.
Making $\mathbf{s}_{1}=\left(  s_{1},...,s_{n}\right)  $ and considering the
geodesic going through $\mathbf{s}_{1}$ and $\mathbf{s}_{1_{j}},$ there is
always an isometry $\varphi$ in $\mathbb{H}^{n}$ which takes this geodesic
into $\mathbb{H},$ characterized as a vertical line. In the case of $n=2,$
$\varphi$ will be a hyperbolic rotation with center in $\mathbf{s}_{1}$ and
radius $d_{\mathbb{H}^{2}}\left(  \mathbf{s}_{1},\mathbf{s}_{1_{j}}\right)  ,$
as illustrated below in Figure 4.

\begin{center}
\includegraphics{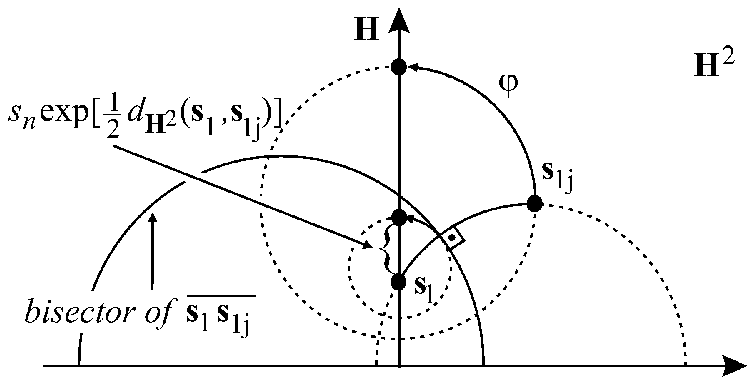}
\end{center}

\begin{center}
\textsc{Figure 4:}\textit{\ Action of the isometry }$\varphi$\textit{\ of
center }$\mathbf{s}_{1}$\textit{\ and radius }$d_{\mathbb{H}^{2}}\left(
\mathbf{s}_{1},\mathbf{s}_{1_{j}}\right)  $\textit{\ in }$\mathbb{H}^{2}.$
\end{center}

\bigskip

Thus, we can write%
\begin{align*}
P\left(  \mathbf{s}_{1},\mathbf{s}_{1_{j}}\right)   &  \leq\int_{s_{n}%
\exp\left(  \frac{1}{2}d_{\mathbb{H}^{n}}\left(  \mathbf{s}_{1},\mathbf{s}%
_{1_{j}}\right)  \right)  }^{+\infty}\dfrac{1}{\sqrt{2\pi\sigma^{2}}}%
\exp\left(  -\dfrac{1}{2\sigma^{2}}\ln^{2}\dfrac{y}{s_{n}}\right)
d_{\mathbb{H}}y\\
&  =\int_{s_{n}\exp\left(  \frac{1}{2}d_{\mathbb{H}^{n}}\left(  \mathbf{s}%
_{1},\mathbf{s}_{1_{j}}\right)  \right)  }^{+\infty}\dfrac{1}{\sqrt{2\pi
\sigma^{2}}}\exp\left(  \ln\left(  \dfrac{y}{s_{n}}\right)  ^{-\frac
{1}{2\sigma^{2}}\ln\frac{y}{s_{n}}}\right)  d_{\mathbb{H}}y\\
&  =\int_{s_{n}\exp\left(  \frac{1}{2}d_{\mathbb{H}^{n}}\left(  \mathbf{s}%
_{1},\mathbf{s}_{1_{j}}\right)  \right)  }^{+\infty}\dfrac{1}{\sqrt{2\pi
\sigma^{2}}}\left(  \dfrac{y}{s_{n}}\right)  ^{-\frac{1}{2\sigma^{2}}\ln
\frac{y}{s_{n}}}d_{\mathbb{H}}y\\
&  =\dfrac{1}{\sqrt{2\pi\sigma^{2}}}\int_{s_{n}\exp\left(  \frac{1}%
{2}d_{\mathbb{H}^{n}}\left(  \mathbf{s}_{1},\mathbf{s}_{1_{j}}\right)
\right)  }^{+\infty}\left(  \dfrac{y}{s_{n}}\right)  ^{\ln\left(  \frac
{y}{s_{n}}\right)  ^{-\frac{1}{2\sigma^{2}}}}\frac{1}{y}dy\\
&  =\dfrac{1}{\sqrt{2\pi\sigma^{2}}}\int_{s_{n}\exp\left(  \frac{1}%
{2}d_{\mathbb{H}^{n}}\left(  \mathbf{s}_{1},\mathbf{s}_{1_{j}}\right)
\right)  }^{+\infty}\dfrac{\left(  y\right)  ^{\ln\left(  \left(  \frac
{y}{s_{n}}\right)  ^{-\frac{1}{2\sigma^{2}}}\right)  -1}}{\left(
s_{n}\right)  ^{\ln\left(  \left(  \frac{y}{s_{n}}\right)  ^{-\frac{1}%
{2\sigma^{2}}}\right)  }}dy\\
&  =\dfrac{1}{\sqrt{2\pi\sigma^{2}}}\int_{s_{n}\exp\left(  \frac{1}%
{2}d_{\mathbb{H}^{n}}\left(  \mathbf{s}_{1},\mathbf{s}_{1_{j}}\right)
\right)  }^{+\infty}\dfrac{\left(  y\right)  ^{\ln\left(  y^{-\frac{1}%
{2\sigma^{2}}}\right)  -\ln\left(  s_{n}^{-\frac{1}{2\sigma^{2}}}\right)  -1}%
}{\left(  s_{n}\right)  ^{\ln\left(  y^{-\frac{1}{2\sigma^{2}}}\right)
-\ln\left(  s_{n}^{-\frac{1}{2\sigma^{2}}}\right)  }}dy\\
&  =\dfrac{\left(  s_{n}\right)  ^{\ln\left(  s_{n}^{-\frac{1}{2\sigma^{2}}%
}\right)  }}{\sqrt{2\pi\sigma^{2}}}\int_{s_{n}\exp\left(  \frac{1}%
{2}d_{\mathbb{H}^{n}}\left(  \mathbf{s}_{1},\mathbf{s}_{1_{j}}\right)
\right)  }^{+\infty}\dfrac{\left(  y\right)  ^{\ln\left(  y^{-\frac{1}%
{2\sigma^{2}}}\right)  -\ln\left(  s_{n}^{-\frac{1}{2\sigma^{2}}}\right)  -1}%
}{\left(  s_{n}\right)  ^{\ln\left(  y^{-\frac{1}{2\sigma^{2}}}\right)  }}dy.
\end{align*}

But $s_{n}>0,$ then $\exists!$ $\mu\in\mathbb{R}$ such that $s_{n}=e^{\mu}.$
Thus,
\begin{align*}
P\left(  \mathbf{s}_{1},\mathbf{s}_{1_{j}}\right)   &  \leq\dfrac{\left(
e^{\mu}\right)  ^{\ln\left(  \left(  e^{\mu}\right)  ^{-\frac{1}{2\sigma^{2}}%
}\right)  }}{\sqrt{2\pi\sigma^{2}}}\int_{\exp\left(  \mu+\frac{1}%
{2}d_{\mathbb{H}^{n}}\left(  \mathbf{s}_{1},\mathbf{s}_{1_{j}}\right)
\right)  }^{+\infty}\dfrac{\left(  y\right)  ^{\ln\left(  y^{-\frac{1}%
{2\sigma^{2}}}\right)  -\ln\left(  \left(  e^{\mu}\right)  ^{-\frac{1}%
{2\sigma^{2}}}\right)  -1}}{\left(  e^{\mu}\right)  ^{\ln\left(  y^{-\frac
{1}{2\sigma^{2}}}\right)  }}dy\\
&  =\dfrac{\left(  e^{\mu_{1}}\right)  ^{-\frac{\mu_{1}}{2\sigma^{2}}}}%
{\sqrt{2\pi\sigma^{2}}}\int_{\exp\left(  \mu+\frac{1}{2}d_{\mathbb{H}^{n}%
}\left(  \mathbf{s}_{1},\mathbf{s}_{1_{j}}\right)  \right)  }^{+\infty}%
\dfrac{\left(  y\right)  ^{\ln\left(  y^{-\frac{1}{2\sigma^{2}}}\right)
+\frac{\mu}{2\sigma^{2}}-1}}{y^{-\frac{^{\mu}}{2\sigma^{2}}}}dy\\
&  =\dfrac{\exp\left(  -\frac{\mu^{2}}{2\sigma^{2}}\right)  }{\sqrt{2\pi
\sigma^{2}}}\int_{\exp\left(  \mu+\frac{1}{2}d_{\mathbb{H}^{n}}\left(
\mathbf{s}_{1},\mathbf{s}_{1_{j}}\right)  \right)  }^{+\infty}y^{\frac{\mu
_{1}}{\sigma^{2}}-1-\frac{1}{2\sigma^{2}}\ln y}dy.
\end{align*}

By Lemma \ref{LemaErfHiperbolico},%
\begin{align*}
&  \dfrac{\exp\left(  -\frac{\mu^{2}}{2\sigma^{2}}\right)  }{\sqrt{2\pi
\sigma^{2}}}\int_{\exp\left(  \mu+\frac{1}{2}d_{\mathbb{H}^{n}}\left(
\mathbf{s}_{1},\mathbf{s}_{1_{j}}\right)  \right)  }^{+\infty}y^{\frac{\mu
_{1}}{\sigma^{2}}-1-\frac{1}{2\sigma^{2}}\ln y}dy\\
&  =\dfrac{\exp\left(  -\frac{\mu^{2}}{2\sigma^{2}}\right)  }{\sqrt{2\pi
\sigma^{2}}}\sqrt{\dfrac{\pi\sigma^{2}}{2}}\exp\left(  \frac{\mu^{2}}%
{2\sigma^{2}}\right)  \operatorname*{erfc}\left(  \frac{d_{\mathbb{H}^{n}%
}\left(  \mathbf{s}_{1},\mathbf{s}_{1_{j}}\right)  }{2\sqrt{2\sigma^{2}}%
}\right) \\
&  =\dfrac{1}{2}\operatorname*{erfc}\left(  \frac{d_{\mathbb{H}^{n}}\left(
\mathbf{s}_{1},\mathbf{s}_{1_{j}}\right)  }{2\sqrt{2\sigma^{2}}}\right)  .
\end{align*}

Consequently,
\[
P\left(  \mathbf{s}_{1},\mathbf{s}_{1_{j}}\right)  \leq\dfrac{1}%
{2}\operatorname*{erfc}\left(  \frac{d_{\mathbb{H}^{n}}\left(  \mathbf{s}%
_{1},\mathbf{s}_{1_{j}}\right)  }{2\sqrt{2\sigma^{2}}}\right)  .
\]

With this result, the probability of transmitting the signal $\mathbf{s}_{1}$
and receiving another one can be upper-bounded as follows%
\[
\mathcal{P}_{e}\left(  \mathbf{s}_{1}\right)  \leq%
{\textstyle\sum\limits_{j=1}^{v_{1}}}
P\left(  \mathbf{s}_{1},\mathbf{s}_{1_{j}}\right)  \leq%
{\textstyle\sum\limits_{j=1}^{v_{1}}}
\dfrac{1}{2}\operatorname*{erfc}\left(  \frac{d_{\mathbb{H}^{n}}\left(
\mathbf{s}_{1},\mathbf{s}_{1_{j}}\right)  }{2\sqrt{2\sigma^{2}}}\right)  .
\]

But since all the signals are \textit{elts},
\[
\mathcal{P}_{e}\leq\dfrac{1}{m}%
{\textstyle\sum\limits_{k=1}^{m}}
\mathcal{P}_{e}\left(  \mathbf{s}_{k}\right)  \leq\dfrac{1}{m}%
{\textstyle\sum\limits_{k=1}^{m}}
{\textstyle\sum\limits_{j=1}^{v_{k}}}
\dfrac{1}{2}\operatorname*{erfc}\left(  \frac{d_{\mathbb{H}^{n}}\left(
\mathbf{s}_{k},\mathbf{s}_{k_{j}}\right)  }{2\sqrt{2\sigma^{2}}}\right)  .
\]

We point out that the second sum is only on the indices of signals that affect
the Voronoi region of $\mathbf{s}_{k}.$\hfill$\square$

\begin{cor}
\label{CorolTeoEstimador}Under the conditions of Theorem
\ref{TeoremaEstimador}, with the additional hypothesis of the set of signals
being geometrically uniform, we have
\[
\mathcal{P}_{e}\leq%
{\textstyle\sum\limits_{j=1}^{v_{1}}}
\dfrac{1}{2}\operatorname*{erfc}\left(  \frac{d_{\mathbb{H}^{n}}\left(
\mathbf{s}_{1},\mathbf{s}_{1_{j}}\right)  }{2\sqrt{2\sigma^{2}}}\right)  .
\]

\end{cor}

It is important to notice the similarity of the hyperbolic bounds with the
Euclidean ones given by Proposition \ref{PropLimitSupGaussRn} and by Corollary
\ref{CorolGeomUnifGaussRn}. Actually, we have just proven that these bounds
are valid for the Euclidean metric as well as for the hyperbolic metric.

We end this section with a comparative analysis between two signals
constellations: the $8$\textit{-PSK} constellation (Figure 1) and the
correspondent one, which will be called $8$\textit{-HPSK}, in hyperbolic plane.

Let $\mathbb{H}^{2}$ the half-plane model. The signals of $8$\textit{-HPSK}
constellation can be obtained by the hyperbolic isometries:%
\[%
\begin{array}
[c]{cccc}%
\mathfrak{e}^{k}: & \mathbb{H}^{2} & \longrightarrow & \mathbb{H}^{2}\\
& \mathbf{z} & \longmapsto & \frac{\mathbf{z}\cos\frac{k\pi}{8}+\sin\frac
{k\pi}{8}}{-\mathbf{z}\sin\frac{k\pi}{8}+\cos\frac{k\pi}{8}}%
\end{array}
,\text{ }k\in\mathbb{N},
\]
where $\mathbb{H}^{2}$ is identified with the complex half-plane. The eight
equidistant points on the unitary hyperbolic circle centred at $\left(
0,1\right)  =i$ are given by $\mathbf{z}=\dfrac{1}{e}i=\left(  0,\dfrac{1}%
{e}\right)  $ and $k=0,...,7$: $\mathbf{s}_{1}=\left(  0\text{ , }0.37\right)
;$ $\mathbf{s}_{2}=\left(  0.35\text{ , }0.42\right)  ;$ $\mathbf{s}%
_{3}=\left(  0.76\text{ , }0.65\right)  ;$ $\mathbf{s}_{4}=\left(  1,17\text{
, }1.40\right)  ;$ $\mathbf{s}_{5}=\left(  0\text{ , }2.72\right)  ;$
$\mathbf{s}_{6}=\left(  -1.17\text{ , }1.40\right)  ;$ $\mathbf{s}_{7}=\left(
-0.76\text{ , }0.65\right)  ;$ $\mathbf{s}_{8}=\left(  -0.35\text{ ,
}0.42\right)  .$

Figure 5 illustrates the $8$\textit{-HPSK} in the half-plane model and the
lines that determinate the Voronoi region of the signals.

\begin{center}
\includegraphics{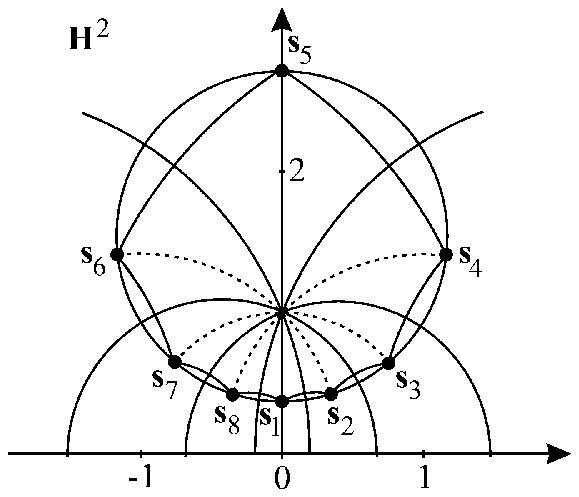}
\end{center}

\begin{center}
\textsc{Figure 5:} $8$\textit{-HPSK} and their Voronoi regions.
\end{center}

\bigskip

This constellation is geometrically uniform in the hyperbolic metric and,
using Corollary \ref{CorolTeoEstimador}, the error probability of any of those
signals satisfies:%
\[
\mathcal{P}_{e}\leq%
{\textstyle\sum\limits_{j=1}^{v_{1}}}
\dfrac{1}{2}\operatorname*{erfc}\left(  \frac{d_{\mathbb{H}^{2}}\left(
\mathbf{s}_{1},\mathbf{s}_{1_{j}}\right)  }{2\sqrt{2\sigma^{2}}}\right)  .
\]
But this sum is over the signal indices that determinate the edges of $V_{1}$
what means it just have two terms, that is:%
\begin{align*}
\mathcal{P}_{e}  &  \leq\dfrac{1}{2}\operatorname*{erfc}\left(  \frac
{d_{\mathbb{H}^{2}}\left(  \mathbf{s}_{1},\mathbf{s}_{2}\right)  }%
{2\sqrt{2\sigma^{2}}}\right)  +\dfrac{1}{2}\operatorname*{erfc}\left(
\frac{d_{\mathbb{H}^{2}}\left(  \mathbf{s}_{1},\mathbf{s}_{8}\right)  }%
{2\sqrt{2\sigma^{2}}}\right) \\
&  =\operatorname*{erfc}\left(  \frac{d_{\mathbb{H}^{2}}\left(  \mathbf{s}%
_{1},\mathbf{s}_{2}\right)  }{2\sqrt{2\sigma^{2}}}\right)  ,
\end{align*}
since $d_{\mathbb{H}^{2}}\left(  \mathbf{s}_{1},\mathbf{s}_{2}\right)
=d_{\mathbb{H}^{2}}\left(  \mathbf{s}_{1},\mathbf{s}_{8}\right)  .$

But%
\begin{align*}
d_{\mathbb{H}^{2}}\left(  \mathbf{s}_{1},\mathbf{s}_{2}\right)   &  =\ln
\dfrac{\left\vert 0.35+0.42i+0.37i\right\vert +\left\vert
0.35+0.42i-0.37i\right\vert }{\left\vert 0.35+0.42i+0.37i\right\vert
-\left\vert 0.35+0.42i-0.37i\right\vert }\\
&  =0.86924
\end{align*}
and then:%
\[
\mathcal{P}_{e}\leq\operatorname*{erfc}\left(  \frac{0.86924}{2\sqrt
{2\sigma^{2}}}\right)  .
\]

Figure 6 illustrates the graph of the upper-bound given above (plotted as
$8$\textit{-HPSK}) compared with the Battacharyya Bound for the standard
$8$\textit{-PSK} in the plane on a circle of radius equal to one.

\begin{center}
\includegraphics{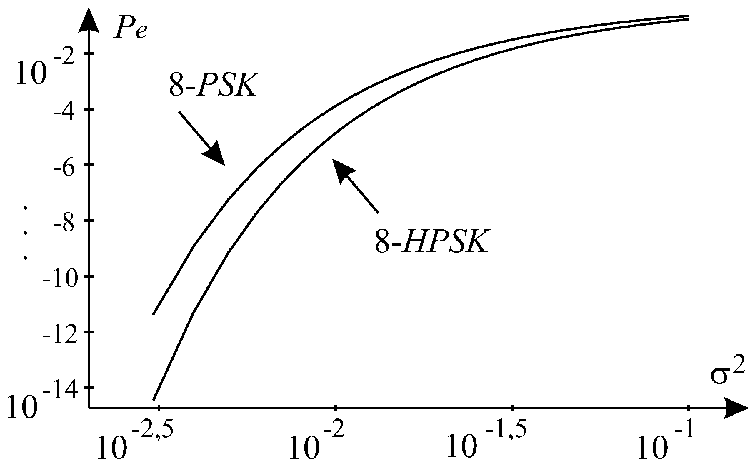}
\end{center}

\begin{center}
\textsc{Figure 6:}\textit{\ Graphs of upper-bounds for error-probability for
}$8$\textit{-HPSK and }$8$\textit{-PSK.}
\end{center}

\bigskip

As we can see both bounds have similar behavior showing a slightly better
performance for error probability in the hyperbolic case. This difference also
increases with the radius of the circles. Such results could somehow be
expected since equidistant points on a hyperbolic circle are farther apart
than their Euclidean counterparts.

\section{\label{Sec Conclud}Concluding Remarks}

The introduction of the concept of a Gaussian probability distribution for the
hyperbolic distance and the derivation of a corresponding upper bound for the
signal transmission error probability presented in this paper provides a tool
for performance comparisons between constellations of points in \ the
hyperbolic environment as well as comparison of these with the usual Euclidean
constellations. This study is then a contribution to considering possible
applications of hyperbolic geometry to coding theory when the signal
transmission can be properly modelled in a $n$-dimensional hyperbolic space.
Some of these new possibilities were pointed out in \cite{Costa},
\cite{EduardoPalazzo2004} and \cite{EduardoArtigo}.

\end{document}